\documentstyle[twocolumn,amsfonts,prl,aps]{revtex}

\newcommand{\beq}{\begin{equation}}
\newcommand{\eeq}{\end{equation}}

\newcommand{\ket}[1]{| #1 \rangle}

\newcommand{\inner}[2]{ \langle #1 | #2 \rangle}
\newcommand{\melement}[3]{ \langle #1 | #2 | #3 \rangle}

\begin{document}
\draft
\title{Quantum Error Correction and Orthogonal Geometry}

\date{July 2, 1996}
\author{A.~R. Calderbank,$^1$ E.~M. Rains,$^2$ P.~W. Shor,$^1$
and N.~J.~A. Sloane$^1$}
\address{$^1$AT\&T Research, Murray Hill, New Jersey, 07974}
\address{$^2$Institute for Defense Analyses, Princeton, New Jersey, 
08540}
\maketitle

\sloppy

\begin{abstract}
A group theoretic framework is introduced that simplifies the 
description of known quantum error-correcting codes and greatly 
facilitates the construction of new examples.  Codes are given 
which map 3 qubits to 8 qubits correcting 1 error, 4 to 10 qubits 
correcting 1 error, 1 to 13 qubits correcting 2 errors, and 
1 to 29 qubits correcting 5 errors.
\end{abstract}
\narrowtext

A quantum error-correcting code is a way of encoding quantum states 
into qubits (two-state quantum systems) so that error or decoherence 
in a small number of individual qubits has little or no effect on the 
encoded data.  The existence of quantum error-correcting codes was 
discovered only recently \cite{Shor}.  These codes have 
the potential to be important for quantum computing and quantum 
communication; for example, they could be used as repeaters for 
quantum cryptography.
Although the subject of quantum error-correcting codes is relatively
new, a large number of 
papers have already appeared.  Many of these
describe specific examples of codes 
\cite{Shor,purify,CS,Steane2,LMPZ,PVK,VGW,SS,BDSW,Braun,Gottes}.  
However, the theoretical aspects of these papers have been
concentrated on properties and rates of the codes 
\cite{BDSW,EM,Schumacher1,Schumacher2,KL,Lloyd}, rather than
on combinatorial recipes for constructing them.
This letter introduces a unifying framework which explains
the codes discovered to date
and greatly facilitates the construction of new examples.

The basis for this unifying framework is group theoretic.  It rests
on the structure of certain subgroups of ${\rm O}(2^n)$ and 
${\rm U}(2^n)$ called {\it Clifford groups} \cite{Kerdock}.
We first construct a subgroup $E$ of ${\rm O}(2^n)$.
This group $E$ is an extraspecial 2-group, and provides a 
bridge between quantum error-correcting codes in Hilbert space and
binary orthogonal geometry.
We obtain the Clifford groups by taking the normalizer $L$
of $E$ in ${\rm O}(2^n)$ or the normalizer $L'$ of the group $E'$ 
generated by $E$ and $iI$ in ${\rm U}(2^n)$.  Since the natural 
setting for quantum mechanics is complex space, it might appear more
appropriate to focus on the complex group $L'$.  However, we shall
begin by discussing the real Clifford group $L$, since
its structure 
is easier to understand, and $L$ is all that is required for
the construction of the known quantum error-correcting codes.

\paragraph*{The Group Theoretic Framework.}
The extraspecial 2-group $E$ is realized as an irreducible 
group of $2^{1+2n}$ orthogonal $2^n \times 2^n$ matrices.  The center
$\Xi(E)$ is $\{ \pm {I} \}$ and the {extraspecial} property is that
$\bar{E} = E/\Xi(E)$ is elementary abelian
(hence a binary vector space).  Let $V$ denote the vector space
${\Bbb Z}_2^{n}$ (where ${\Bbb Z}_2=\{0,1\}$)
and label the standard basis of 
${\Bbb R}^{2^n}$ by
$\ket{v}$, $v \in V$.  Every element $e$ of $E$ can be written
uniquely in the form
\begin{equation}
e = X(a)Z(b) (-{I})^\lambda
\end{equation}
where $\lambda \in {\Bbb Z}_2$, $X(a): \ket{v} \rightarrow \ket{v+a}$,
$Z(b): \ket{v} \rightarrow (-1)^{b\cdot v} \ket{v}$, 
for $a$, $b \in V$.

Consider the $j$th qubit in a quantum channel which 
transmits $n$ qubits.  Let $v_j$ be
the vector with a 1 in the $j$th bit and 0's in the 
remaining bits.  Then $X(v_j)$ is the transformation which 
\renewcommand{\arraystretch}{.33}
applies the Pauli matrix $\sigma_x =
\left(\begin{array}{cc} \scriptstyle 0& \scriptstyle 1
\\ \scriptstyle 1 & \scriptstyle 0\end{array}\right)$
to the $j$th qubit and does nothing to the remaining $n-1$ qubits. 
The transformation $Z(v_j)$ 
applies the Pauli matrix $\sigma_z =
\left(\begin{array}{cc} \scriptstyle 1& \scriptstyle 0
\\ \scriptstyle 0 & \scriptstyle -1\end{array}\right)$
\renewcommand{\arraystretch}{1}
to the $j$th bit and does nothing to the other $n-1$ qubits.  In 
the language of quantum error correction,
$X(v_j)$ is a bit error and $Z(v_j)$ is a
phase error in the $j$th qubit.  The element $X(a)Z(b)$ corresponds
to bit errors in the qubits for which $a_j=1$ and phase
errors in the qubits for which $b_j = 1$.

Observe that $E$ is the group of tensor products $\pm w_1 \otimes
\ldots \otimes w_n$ where each $w_j$ is one of $\sigma_x$, $\sigma_z$
or $\sigma_y=\sigma_x\sigma_z$. 
For the purposes of quantum error correction, we need consider
only errors of the types $\sigma_x$, $\sigma_z$, and 
$\sigma_y$, since if we can correct
these errors we can correct arbitrary 
errors \cite{Steane2,BDSW,EM}.

Define a quadratic form $Q$ on $\bar{E} = E/\Xi(E)$ by
\begin{equation}
Q(\bar{e}) = \sum_{j=0}^n a_j b_j,
\end{equation}
where $e = \pm X(a) Z(b)$ is any element of $E$ whose image
in $\bar{E}$ is $\bar{e}$. 
Then $e^2 = (-I)^{Q(\bar{e})}$ and 
$Q(\bar{e}) = 0$ or $1$ according as 
$X(a)$ and $Z(b)$ commute or anticommute.
If $e = w_1 \otimes \ldots \otimes w_n$ then 
$Q(\bar{e})$ is the parity of the number of components $w_j$ that
are equal to $\sigma_x \sigma_z$.  

The normalizer $L$ of $E$ in the real orthogonal group $O({\Bbb R}^{2^n})$
(the subgroup of elements $g$ such that $g^{-1} E g = E$)
acts on $E$ by conjugation, fixing the center $\Xi(E)$ 
($g \in L$ acts on $E$ as the permutation $e \rightarrow g^{-1}eg$).
Hence there is a well-defined action of $L$ on the binary vector space
$\bar{E}$ that preserves the quadratic form $Q$.  The quotient $L/E$ is
the orthogonal group ${\rm O^+}(2n,2)$, a finite classical group
\cite{group-theory}.  The group $L$ appears in recent connections
between classical Kerdock error-correcting codes, orthogonal geometry,
and extremal Euclidean line sets~\cite{Kerdock}.  This group also
appears \cite{BDSW} as the group of Bell-state-preserving bilateral
local transformations that two experimenters ($A$ and $B$) can 
jointly perform
on $n$ pairs of particles (each pair being in a Bell state).
Hence there is a one-to-one correspondence between Bell states
and elements of $\bar{E}$ (cf.\ Eqs.~(39) and (67) of \cite{BDSW}).  
The quadratic form $Q(\bar{e})$ is $0$ or $1$ according as the Bell states
are symmetric or antisymmetric under interchange of $A$ and $B$.  
Observe that symmetry/antisymmetry of Bell states is the physical
invariant conserved by this presentation of $L$.

The following are group elements that generate $L$, shown together
with their induced action
on the binary vector space $\bar{E}$.

(1) $H = 2^{-n/2} [ (-1)^{u\cdot v} ] _{u,v \in V}$,
which interchanges $X(b)$ and $Z(b)$.

(2) Every matrix $A$ in the general linear group ${\rm GL}(V)$
determines a permutation matrix $\ket{v} \rightarrow \ket{vA}$ in 
${\rm O}({\Bbb R}^{2^n})$.  The action on $\bar{E}$ induced by
conjugation is $X(a) \rightarrow X(aA)$, 
$Z(b) \rightarrow Z(bA^{-{\rm T}})$.  For example the quantum {\sc xor}
taking $\ket{q_1q_2} \rightarrow \ket{q_1 (q_1 \oplus q_2)}$ is
represented by
\begin{equation}
(a_1 a_2 | b_1 b_2) \rightarrow (a_1 a_2 | b_1 b_2) 
\left(
\begin{array}{cc|cc}
1&1&0&0 \\
0&1&0&0 \\
\hline
0&0&1&0 \\
0&0&1&1 
\end{array}
\right).
\end{equation}
The back action of the {\sc xor} on the phases is evident in 
its effect on $b_1$ and $b_2$.  Any orthogonal matrix in 
${\rm O}({\Bbb R}^{2^n})$ that normalizes both $X(V)$ 
and $Z(V)$ is of this type for some $A$
\cite[Lemma 3.14]{Kerdock}.

(3) $\hat{H}_2 = \frac{1}{\sqrt{2}}
\renewcommand{\arraystretch}{.33}
\left(\begin{array}{cc} \scriptstyle 1& \scriptstyle 1
\\ \scriptstyle 1 & \scriptstyle -1\end{array}\right)
\renewcommand{\arraystretch}{1}
\otimes {I}_{2^{n-1}}$, which applies a $\pi/2$ rotation to 
the first qubit and leaves the other qubits unchanged.  
This acts on ${\bar{E}}$ by interchanging
$a_1$ and $b_1$.

(4) Diagonal matrices $d_M = {\rm diag}[(-1)^{Q_M(v)}]$, where
$Q_M$ is a binary quadratic form on $V$ for which the associated 
bilinear form $Q_M(u+v)-Q_M(u) - Q_M(v)$ is $u M v^{\rm T}$.  
Note that $M$ is symmetric with zero diagonal.  The induced action on 
$\bar{E}$ is given by 
\begin{equation}
(a|b) \rightarrow (a|b) 
\left(
\begin{array}{c|c}
{I} & M\\
\hline
0 & {I}
\end{array}
\right).
\end{equation}
These matrices are precisely the elements of $L$ that induce the 
identity on the
subgroup $Z(V)$.  In terms of their effect on qubits, these are the
transformations in $L$ that change the phases of the qubits while
fixing their values.

{\it Remark.}  The group $L'$ is the normalizer of the group $E'$ 
generated by $E$ and $i{I}$ in the unitary group ${\rm U}(2^n)$.
Now we cannot define $Q(\bar{e}) = e^2$ because $(ie)^2 \neq
e^2$.  However, we still have the nonsingular alternating binary
form
\begin{equation}
\left(\bar{X}(a) \bar{Z}(b), \bar{X}(a') \bar{Z}(b')\right)
= a\cdot b' + a' \cdot b.
\end{equation}
The group $L'$ is generated by $L$ and by diagonal transformations 
$d_P = {\rm diag}[i^{T_P(v)}]$ where $T_P$ is a ${\Bbb Z}_4$-valued
quadratic form \cite[Section 4]{Kerdock}.  The induced action on 
$\bar{E}' = E' / \Xi(E')$ is described by 
\begin{equation}
(a|b) \rightarrow (a|b)
\left(
\begin{array}{c|c}
{I} & P\\
\hline
0 & {I}
\end{array}
\right)
\end{equation}
where $P$ is symmetric and may have a nonzero diagonal.
For example, applying the transformation
\renewcommand{\arraystretch}{.33}
$\left(\begin{array}{cc} \scriptstyle 1& \scriptstyle 0
\\ \scriptstyle 0 & \scriptstyle i\end{array}\right)$
\renewcommand{\arraystretch}{1}
to each qubit corresponds to $P={I}$.  The
quotient $L'/E'$ is the symplectic group ${\rm Sp}(2n,2)$
\cite{group-theory}.

\paragraph*{Quantum Error-Correcting Codes.}
A subspace $\bar{S}$ of $\bar{E}$ is said to be totally singular if
$Q(\bar{s}) = 0$ for all $\bar{s} \in \bar{S}$.  It follows that for $\bar{s} = (a|b)$, 
$\bar{s}' = (a'|b')$ in $\bar{S}$ the inner product 
$(\bar{s},\bar{s}') = Q(\bar{s}+\bar{s}') - Q(\bar{s}) - Q(\bar{s}')$ satisfies 
\begin{equation}
(\bar{s},\bar{s}') = a\cdot b' + a' \cdot b = 0.
\label{inner-product}
\end{equation}
The group $L$ acts transitively on totally singular subspaces of a
given dimension.  Hence every $k$-dimensional totally singular subspace
is contained in the same number of maximal ($n$-dimensional) totally
singular subspaces.  If $\bar{M}$ is a maximal totally singular 
subspace, then the group $\bar{M}$ has $2^n$ distinct linear characters, 
and the corresponding eigenspaces determine a coordinate frame 
${\cal F}(\bar{M})$ (an orthonormal basis of ${\Bbb R}^{2^n}$).  
For example, the image of $Z(V)$ in $\bar{E}$ determines
the coordinate frame $\ket{v}$, $v \in V$, and the image
of $X(V)$ determines the
coordinate frame $2^{-n/2} \sum_v (-1)^{u \cdot v} \ket{v}$, $u \in V$.
If $\bar{S} \subseteq \bar{M}$ is a $k$-dimensional totally singular
subspace, then the group $\bar{S}$ has $2^k$ distinct linear 
characters.  The
$2^n$ vectors in ${\cal F}(\bar{M})$ are partitioned into $2^k$ sets of 
size $2^{n-k}$ with each set corresponding to a different eigenspace.
We view each eigenspace as a quantum error-correcting code
which maps $n-k$ qubits into $n$ qubits.
The $2^{n-k}$ vectors from ${\cal F}(\bar{M})$ in that eigenspace constitute
the codewords.

More generally, we may use complex space, and so we define a 
quantum error-correcting code encoding $k$ qubits
into $n$ qubits to be any $2^k$-dimensional 
subspace $C$ of ${\Bbb C}^{2^n}$.
This code will protect against errors in a certain
error set ${\cal E}$, which we
take to be a subset of the extraspecial group $E$.
As remarked earlier, there is no loss of generality
in restricting to error sets in $E$.

For such a code $C$ to protect against all errors 
in an error set $\cal E$, it is necessary and sufficient 
\cite{BDSW,KL} that
for any two vectors $\ket{c_1}$ and $\ket{c_2}$ in $C$ with
$\inner{c_1}{c_2} = 0$, and any two transformations $e_1$ and $e_2$
from ${\cal E}$, we have
\begin{eqnarray}
\melement {c_1}{e_1^{-1} e_2 }{c_2} &=& 0,
\label{first-cond}
\\
\melement {c_1}{e_1^{-1} e_2 }{c_1} &=&
\melement {c_2}{e_1^{-1} e_2 }{c_2}.
\label{second-cond}
\end{eqnarray}
Note that since we are assuming that
${\cal E} $ is contained in $E$, we can replace $e_1^{-1} e_2$ by
$e_1 e_2$ in the above equations, since
$e_1 = \pm e_1^{-1}$.
An interesting special case occurs when both sides of 
Eq.~(\ref{second-cond})
are always equal to 0 for $e_1 \neq e_2 \in {\cal E}$.  This implies 
that there is a measurement which will uniquely determine the error 
without affecting the encoded subspace.  After this measurement, the 
error can subsequently be corrected by a unitary operation.  

If both sides of Eq.~(\ref{second-cond}) are not always 0, 
then there can be two errors $e_1$ and $e_2$ between which it is
impossible to distinguish.  However, these two errors are guaranteed
to have identical effects on vectors within the subspace $C$, and so
it is not necessary to distinguish between these errors in order to
correct the error.

We now can show the connection between orthogonal geometry and quantum 
error correcting codes.

{\it Theorem 1.}  Suppose that $\bar{S}$ is a $k$-dimensional
totally singular subspace of $\bar{E}$.  Let $\bar{S}^\perp$ be the 
$(2n-k)$-dimensional subspace orthogonal to $\bar{S}$ with 
respect to the inner product (\ref{inner-product}).  Further 
suppose that for any two vectors ${e}_1$ and ${e}_2$ in 
an error set ${\cal E} \subseteq {E}$, 
either $\bar{e}_1 \bar{e}_2 \in \bar{S}$ or $\bar{e}_1 \bar{e}_2 
\notin \bar{S}^\perp$.  Then
the eigenspace $C$ corresponding to any character of the group 
$\bar{S}$ is an error-correcting code which will correct any error 
$e \in {\cal E}$.

{\it Proof.}  We first show that if $e \in E$, then $e$ permutes the
$2^{k}$ spaces $C_i$ which are generated by the $2^k$ different linear
characters of $\bar{S}$.  
Consider an element $\bar{s} \in \bar{S}$ with eigenvalue $\lambda_s$.
We write $s$ for the associated representation.
Then for any $\ket{c} \in C$  we have $s\ket{c} = \lambda_s\ket{c}$,
and $s e \ket{c} = (-1)^{(\bar{s},\bar{e})} e s \ket{c} 
= (-1)^{(\bar{s},\bar{e})} \lambda_s e \ket{c}$, where 
$(\bar{s},\bar{e})$ is the inner product (\ref{inner-product}).  Since 
$ (-1)^{(\bar{s},\bar{e})} \lambda_s $ is independent of $c$, this shows
that the action of $e$ permutes the eigenspaces generated by the 
characters of $S$.

We will divide the proof into two cases, according as
$\bar{e}_1 \bar{e}_2 \in \bar{S}$ or $\bar{e}_1 \bar{e}_2 
\not\in \bar{S}^\perp$.  

{\it Case 1.}  Suppose $\bar{e}_1 \bar{e}_2 \in \bar{S}$.  It 
follows that
for $\ket{c_1}$ and $\ket{c_2} \in C$ with $\inner{c_1}{c_2} = 0$, 
\begin{equation}
\melement{c_1}{e_1 e_2}{c_2} = \lambda_{e_1 e_2} \inner{c_1}{c_2} = 0 ,
\end{equation}
and for all $\ket{c} \in C$,
\begin{equation}
\melement{c}{e_1 e_2}{c} = \lambda_{e_1e_2}\inner{c}{c} = 
\lambda_{e_1e_2},
\end{equation}
establishing Eqs.~(\ref{first-cond}) and
(\ref{second-cond}).  

{\it Case 2.}  Suppose $\bar{e}_1 \bar{e}_2 \notin \bar{S}^\perp$.  
It follows that for some $\bar{s} \in \bar{S}$, $s e_1 e_2 = - e_1 e_2 s$. Thus, 
for $\ket{c} \in C$, $s e_1 e_2 \ket{c} = - e_1 e_2 s \ket{c} =
- \lambda_s e_1 e_2 \ket{c}$, so $e_1 e_2 \ket{c} \notin C$.  Thus
$e_1 e_2 \ket{c}$ is in a different eigenspace, so
\begin{equation}
\melement{c_1}{e_1e_2}{c_2} = 0
\end{equation}
for all $c_1$, $c_2 \in C$ (including $c_1 = c_2$).  
Again Eqs.~(\ref{first-cond}) and
(\ref{second-cond}) hold.  
\hfill $\Box$

The quadratic form $Q$ plays no role in this proof and it is only
necessary that $\bar{S}$ satisfy $(\bar{S},\bar{S})=0$ with
respect to the alternating form.  This means the complex group $L'$ can
also be used for code construction.

The group $L$ acts transitively on the totally singular subspaces 
$\bar{S}$ of dimension $k$,
so there is some element
$g \in L$ taking any particular $k$-dimensional totally singular 
subspace to the subspace corresponding to a quantum code generated
as in Theorem 1.  This implies that some $g$ takes the canonical 
$2^{n-k}$-dimensional
Hilbert space generated by the first $n-k$ qubits to the
encoded subspace.  Since $L$ can be generated by {\sc xor}'s and $\pi/2$
rotations, these quantum gates are therefore
sufficient for encoding any of 
these quantum codes.

\paragraph*{Example 1.}
We first describe the
code mapping 1 qubit into 5 qubits presented in 
Ref.~\cite{BDSW}, which contains two codewords, 
\begin{eqnarray}
\ket{c_0} &=& \ket{00000} \nonumber \\
&+& \ket{11000} +\ket{01100} +\ket{00110} +\ket{00011} +\ket{10001}  
\nonumber \\
&-& \ket{10100} -\ket{01010} -\ket{00101} -\ket{10010} -\ket{01001}  
\nonumber \\
&-& \ket{11110} -\ket{01111} -\ket{10111} -\ket{11011} -\ket{11101} 
\nonumber , \\
\ket{c_1} &=& \ket{11111} \nonumber  \\
&+& \ket{00111} +\ket{10011} +\ket{11001} +\ket{11100} +\ket{01110}  
\nonumber \\
&-& \ket{01011} -\ket{10101} -\ket{11010} -\ket{01101} -\ket{10110}  
\nonumber \\
&-& \ket{00001} -\ket{10000} -\ket{01000} -\ket{00100} -\ket{00010} 
\nonumber .
\end{eqnarray}
Essentially the same code is given in Ref.~\cite{LMPZ}, but we use the
above presentation since it is fixed under cyclic permutations.
It is easily verified that $X(11000) Z(00101)$ fixes 
$\ket{c_0}$ and $\ket{c_1}$.
Thus we may take 
the vector $(11000|00101) \in \bar{E}$ to be in
the subspace $\bar{S}$.  Using the fact that
the code is closed under cyclic
permutations, we find that $\bar{S}$ is 
a 4-dimensional totally singular subspace 
generated by the vectors
\begin{equation}
\begin{array}{c|c}
11000 & 00101 \\
01100 & 10010 \\
00110 & 01001 \\
00011 & 10100
\end{array} 
\label{five-bit}
\end{equation}
[The fifth cyclic shift, $(10001|01010)$, is also in this subspace.]  
The dual $\bar{S}^\perp$ 
is generated by $\bar{S}$ and the additional vectors 
$(11111|00000)$ and $(00000|11111)$.  It is straightforward to verify
that the minimal weight vectors in $\bar{S}^\perp$ have weight three
[one such vector is $(00111|00101)$] and thus the code can
correct one error.

\paragraph*{Example 2.} Suppose we have a classical linear $[n,k,d]$
binary error correcting code $C$ 
(i.e., it is over ${\Bbb Z}_2^n$, it is 
$k$-dimensional, and it has minimal distance $d$, so that it 
corrects $t=[(d-1)/2]$ errors).  Suppose furthermore that 
$C^\perp \subset C$.  We can define a subspace $\bar{S}$ to consist of
all vectors $(v_1|v_2)\in \bar{E}$, where $v_1$, $v_2 \in C^\perp$.  
The dual $\bar{S}^\perp$ consists of all vectors 
$(v_1|v_2)$ with $v_1$, $v_2 \in C$, showing that the corresponding
quantum error-correcting code corrects $t$ errors. 
The subspace $\bar{S}$ is $2(n-k)$-dimensional, so the quantum code
maps $n-2k$ qubits into $n$ qubits.  This is the method described in
Refs.~\cite{CS,Steane2}.

\paragraph*{Example 3.} Consider the subspace $\bar{S}$ obtained by modifying the 
classical [8,4,4] Hamming code as follows: 
\begin{equation}
\begin{array}{c|c}
01110100 & 00111010 \\
00111010 & 00011101 \\
00011101 & 01001110 \\
11111111 & 00000000 \\
00000000 & 11111111.
\end{array} 
\end{equation}
It is straightforward to verify that these vectors generate a 
5-dimensional totally singular subspace $\bar{S}$ which is invariant
under cyclic permutations of the last 7 bits, and that $\bar{S}^\perp$
has minimal weight 3.  This gives a quantum code
mapping 3 qubits into 8 qubits which can correct one error.  The same
code was discovered by Gottesman \cite{Gottes}, 
also via group-theoretic techniques.

\paragraph*{Example 4.} By duplicating the 5-qubit code (\ref{five-bit}) 
and adding two vectors,
we can obtain a code which maps 4 qubits into
10 qubits and corrects one error:  
\begin{equation}
\begin{array}{cc|cc}
01100 & 11110 & 10010 & 01100 \\
00110 & 01111 & 01001 & 00110 \\
00011 & 10111 & 10100 & 00011 \\
10001 & 11011 & 01010 & 10001 \\
11111 & 11111 & 00000 & 00000 \\
00000 & 00000 & 11111 & 11111 .
\end{array} 
\end{equation}

\paragraph*{Example 5.} The following construction is a generalization
of the 5-qubit code (\ref{five-bit})
inspired by classical quadratic residue codes.
It works for any prime $p$ of the form $8j+5$.  We have not found good
theoretical bounds on the minimal distance, but for small primes
these codes are excellent.  To construct the first vector $(a|b)$, 
put $a_j = 1$ when $j$ is a nonzero quadratic residue mod~$p$ (that is, 
$j = k^2$ mod~$p$ for some $k$) and put $b_j = 1$ when $j$ is a
quadratic nonresidue.  To obtain $p-1$ vectors that generate
the subspace $\bar{S}$, take $p-2$ cyclic shifts of the first vector.
For $p=13$, the first basis vector is
\begin{displaymath}
0101100001101|0010011110010 
\end{displaymath}
and the remaining vectors are obtained by 
cyclic shifts.  The minimal weight of $\bar{S}^\perp$ was calculated
by computer to be 5.  This gives a code mapping one qubit into
13 qubits which corrects 2 errors.
For $p=29$, ths subspace $\bar{S}^\perp$ has minimal distance 
11, so this construction gives a code mapping 1 qubit to 29 
which corrects 5 errors.

We next give a Gilbert--Varshamov lower bound for the asymptotic rate 
of this class of codes.  It matches the Gilbert--Varshamov lower bounds 
known for general quantum error-correcting codes \cite{EM}.

{\it Theorem 2.}  There exist quantum error-correcting codes with
asymptotic rate
\begin{equation}
R = 1 - 2\delta \log_2 3 - H_2 ( 2\delta )
\end{equation}
where $\delta$ is the fraction of qubits that are subject to 
decoherence and $H_2(\delta) = -\delta \log_2 \delta - (1-\delta)
 \log_2 (1- \delta) $ is the binary entropy function.

{\it Proof.}  Let $N_k$ denote the number of $k$-dimensional totally
singular subspaces.  We count pairs $(e,\bar{S})$ where $e \in E$ is 
in ${\cal E}^2$ (i.e., $e = e_1e_2$ with $e_1$, $e_2$ in the error 
set $\cal E$) and $\bar{S}$ is a $k$-dimensional totally singular
subspace with $\bar{e} \in \bar{S}^\perp \backslash \bar{S}$.  
Transitivity of $L$ on singular points ($\bar{e} \neq 0$, 
$Q(\bar{e}) = 0$), and on nonsingular
points ($Q(\bar{e}) \neq 0$) implies that each $e \in E$ satisfies 
$\bar{e} \in \bar{S}^\perp \backslash \bar{S}$ for $\mu N_k$ subspaces 
$\bar{S}$, where the fraction $\mu \approx 2^{-k}$.  If 
$|{\cal E}^2| < 2^k$ then there
exists a $k$-dimensional totally singular subspace $\bar{S}$ that 
satisfies $\bar{e} \notin \bar{S}^\perp \backslash \bar{S}$ for all 
$e \in {\cal E}^2$.  Hence the achievable rate $R$ satisfies
\begin{eqnarray}
\textstyle 1-R &=& \log_2 |{\cal E}^2 | /n \nonumber \\
&=& \log_2 \left[{\textstyle 3^{2\delta n} {n \choose 2\delta n}} \right] /n \\
&=& \textstyle 2\delta \log_2 3 + H_2(2\delta).  
\nonumber \ \ \ \ \ \ \ \ \ \ \ \  \Box
\end{eqnarray}

We would like to thank David DiVincenzo for
discussions about the group $L$ as presented in \cite{BDSW}.

\end{document}